\documentclass[11pt]{article}
\usepackage{jcapmod}

\usepackage{graphicx,amsmath,amssymb}
\usepackage{jcapmod}

\usepackage{booktabs}
\usepackage[english]{babel}
\usepackage{amsmath,amssymb,amsbsy,amstext, amsthm}
\usepackage{hyperref}
\usepackage{graphicx}
\usepackage{amsfonts}
\usepackage{amssymb}
\usepackage[small]{caption}
\usepackage{upgreek}
\usepackage{framed}
\usepackage{wrapfig}
\usepackage{multirow}
\usepackage{subfig}
\usepackage{bbm}
\usepackage{bm}

\usepackage{colortbl}
\definecolor{lightgray}{gray}{0.85}
\definecolor{lightgray2}{gray}{0.9}
\definecolor{lightgreen}{cmyk}{0.2, 0, 0.2, 0.2}
\definecolor{lightgray}{cmyk}{0.1,0.2,0,0.1}
\definecolor{lightgray2}{cmyk}{0.1,0.1,0,0.1}

\makeatletter
\newlength{\apb@width}
\newcommand{\autoparbox}[2][c]{\settowidth{\apb@width}{#2}\parbox[#1]{\apb@width}{#2}}

\makeatother

\numberwithin{equation}{section}

\def\beq{\begin{equation}}
\def\eeq{\end{equation}}
\def\be{\begin{eqnarray}}
\def\ee{\end{eqnarray}}
\def\beq{\begin{equation}}
\def\eeq{\end{equation}}
\def\bea{\begin{eqnarray}}
\def\eea{\end{eqnarray}}

\def\Beq{\begin{equation}\begin{aligned}}
\def\Eeq{\end{aligned}\end{equation}}

\def\dd{{\rm d}}
\def\Tr{{\rm Tr}}

\def\dd{{\rm d}}
\def\Tr{{\rm Tr}}

\DeclareRobustCommand{\SkipTocEntry}[4]{}

\newcommand{\reading}[1]{\hfill $ _{\text{\hyperref[#1]{ref}}} $}

\begin{document}
\begin{titlepage}

\setcounter{page}{1} \baselineskip=15.5pt \thispagestyle{empty}

\bigskip\

\bigskip
\vspace{1cm}
\begin{center}
{\fontsize{20}{24}\selectfont  \sffamily \bfseries  Vacua on the Brink of Decay}
\end{center}
%\title{Vacua on the Brink of Decay}

\vspace{0.2cm}
\begin{center}
{\fontsize{13}{30}\selectfont  Guilherme L. Pimentel$^{1}$, Alexander M. Polyakov$^{2}$ and Grigory M. Tarnopolsky$^{3}$} 
\end{center}

\begin{center}

\vskip 8pt
\textsl{$^1$  Institute of Theoretical Physics, University of Amsterdam,\\Science Park 904, Amsterdam, 1098 XH, The Netherlands}

\vskip 8pt
\textsl{$^{2}$ Joseph Henry Laboratories, Princeton University, \\Princeton, NJ 08544, USA}  

\vskip 8pt
\textsl{$^{3}$ Department of Physics, Harvard University,\\ Cambridge, MA 02138, USA}  
\vskip 7pt
\end{center}

\vspace{1.2cm}
\hrule \vspace{0.3cm}
\noindent {\sffamily \bfseries Abstract} \\[0.1cm]
We consider free massive matter fields in static scalar, electric and gravitational backgrounds. Tuning these backgrounds to the brink of vacuum decay, we identify a term in their effective action that is singular. This singular term is universal, being  independent of the features of the background configuration. In the case of gravitational backgrounds, it can be interpreted as a quantum mechanical analog of Choptuik scaling. If the background is tuned slightly above the instability threshold, this singular term gives the leading contribution to the vacuum decay rate.

\vskip 10pt
\hrule
\vskip 10pt

\vspace{4cm}

\begin{center}
{\it Dedicated to the memory of Ludwig Faddeev}
\end{center}
 \end{titlepage}

\tableofcontents

\newpage
%=============================
\section{Introduction}
%=============================
%%%%%%%%%%%%%%%%%%%%%%%%%%%%%%%%%%%%%%%%%%
In this article, we study strong background fields which may be able to destroy their own environment. This happens when the mass gap of the Quantum Field Theory (QFT) in question, due to the external field, tends to zero and eventually becomes negative. We identify a universal singularity in the effective action of the background field, which signals instability of the vacuum, as the mass gap vanishes.

Background field configurations which lead to particle production are associated to the formation of a ``horizon", i.e., a length scale in which it becomes energetically more favorable to produce particles than to sustain the field configuration. This definition of the horizon is more general than the one usually discussed in the literature. For example, it implies the existence of electromagnetic horizons. As a typical example of an electric horizon, consider electrically charged particles of mass $m$ in a background electrostatic potential $a_t(x)$. It is clear that if the ``voltage" $A\equiv a_t(+\infty)-a_t(-\infty)$ satisfies $A>2 m$ particles will be  produced and not much will remain of the vacuum\footnote{Similar considerations apply to magnetically charged particles in a background magnetostatic field.}. For a gravitational background, the location of the horizon defined in this new way agrees with that of the causal horizon (see e.g. \cite{Wald:1984rg}). 

Technically, the phenomenom of particle production can be diagnosed by calculating the vacuum decay rate, given by the imaginary part of the one-loop effective action, obtained after integrating out  massive matter fields. Many results are available for the effective action for a constant background field; for an incomplete list of references, see \cite{Schwinger:1951nm, Parker:1969au,Hawking:1974sw,Gibbons:1977mu, Savvidy:1977as,Nielsen:1978rm,Polyakov:2007mm} and \cite{Birrell:1982ix,Dunne:2004nc} for reviews. In these examples, the horizon is always present\footnote{The critical field strength associated to Schwinger $e^+e^-$ pair production, $E_c=m_e^2/e$, gives the electric field value for which the pair production rate becomes non-exponentially suppressed. The pair production rate is, however, nonzero for any value of the background constant electric field, $\Gamma\sim (e^2 E)\exp(-E_c/E)$.}. In order to study particle production for field configurations near the threshold, we must consider a gapped matter sector coupled to background fields. The mass gap acts as a barrier, preventing particle production for weak backgrounds. We would like to find singular terms  in the effective action  as we approach the particle production threshold. In this regime, the effective action is real. If we dial the background field strength above the threshold, the effective action acquires an imaginary piece coming from the singular term. This imaginary part gives the vacuum decay rate. 

In this article, we consider different scenarios of strong background fields. We consider background scalar, electric and gravitational fields. We couple these backgrounds to free massive scalar matter, and determine the singular terms in the one-loop determinant of the matter fields in the background geometry tuned to the vicinity of the threshold. The fixed backgrounds are not necessarily a solution of the source-free equations of motion; we assume that there are suitable sources that sustain the static background configuration, and focus on the quantum mechanical response of matter fields to the background.

The electromagnetic threshold singularity might be experimentally testable in the near future by producing strong electric pulses with lasers. The gravitational threshold singularity is a quantum analogue of Choptuik scaling~\cite{Choptuik:1992jv}. Choptuik numerically simulated the gravitational collapse of a distribution of dust particles. If the initial data is tuned above a critical value, the final state has a black hole. Choptuik recognized a remarkable scaling law in the mass of the black hole, as a function of how much above criticality the initial data is. For various setups of initial data, he found scaling laws for the black hole mass with the same exponent. Our result shares the same robustness to the shape of the gravitational potential.

While the exact result for the effective action depends on the details of the background field, we argue that the threshold singularity is universal. The physical reason for that is the following: right above threshold, the first pair production event will be very soft, and the pair will have a very long wavelength. As the wavelength of the excitation is very long, it is not sensitive to the fine features of the background. 

For an early discussion of these ideas, see \cite{PiTPlec}. For recent work that partially overlaps with the results presented here, see \cite{Gies:2016coz}, where universality of the particle production rate is found for electric fields slightly above threshold. Our below-threshold singularity, when extrapolated above the threshold, agrees with the result reported in \cite{Gies:2016coz}. 

\paragraph{Outline}  In section \ref{s:threshold}, we compute the threshold singularity for three different backgrounds -- scalar, electric and gravitational fields. In section \ref{s:conclusions}, we present our conclusions. In appendix \ref{app:a}, we derive in detail a formula for the gravitational effective action in terms of the transmission coefficient; all other cases  follow a similar derivation. In appendix \ref{app:b}, we discuss the behavior of the transmission coefficient when the mass gap is small. Finally, in appendix \ref{app:c}, we quote the exact transmission coefficients for some electric and gravitational backgrounds. The exact results agree with our general considerations in the main text.

\section{Threshold Singularities}\label{s:threshold}

%With the formalism at hand, and an expression \eqref{EAfqm} that relates the effective action to scattering data, 

Having set the stage, let us study the threshold singularity for various quantum field theories in background fields.  In this section, we determine the piece in the effective action which becomes singular as a parameter in the external field configuration reaches the threshold. At the threshold, a very long wavelength pair is produced, which can only probe the rough features of the external field configuration. This allows us to find a universal answer for the singularity, regardless of the precise shape of the external field. The nature of the singularity is slightly different for scalar, vector and gravitational external fields.

\subsection{Scalar Fields}

\begin{figure}[h!]
\centering
\begin{tabular}{cc}
\includegraphics[width=0.48\textwidth]{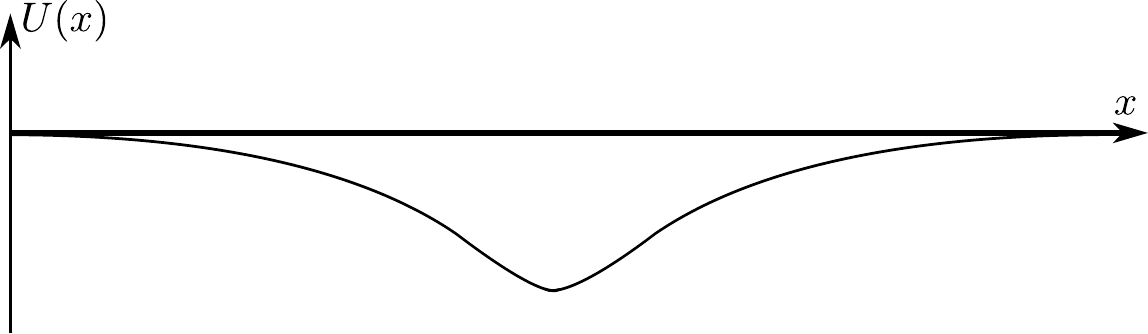}
\end{tabular}
\caption{\label{scalarbackp1}  Plot of a schematic form of  the potential $U(x)$. As its parameters are tuned, the potential becomes deeper. At threshold, the background is too strong, and a bound state of the quantum field is spontaneously produced.}
\end{figure}

We first consider a $1+1$  quantum field theory of a free massive scalar field in a static, position dependent background  $U(x)$. We want to compute the one-loop effective action
\beq
\langle {\rm out} | {\rm in}\rangle = e^{i S_{\rm eff}(U)}=\int D\phi \, \exp\left(\frac{i}{2} \int \dd t \dd x \left((\partial_t\phi)^{2}-(\partial_x\phi)^2-(m^2+U(x))\phi^2\right)\right)\,,
\eeq
where we take $U(x)$ to be an arbitrary smooth function with asymptotic values $U(\pm\infty)\to 0$ (see figure \ref{scalarbackp1}). A very similar model was studied from a different point of view in \cite{migdal1978fermions}.

\begin{figure}[h!]
\centering
\begin{tabular}{cc}
\includegraphics[width=0.48\textwidth]{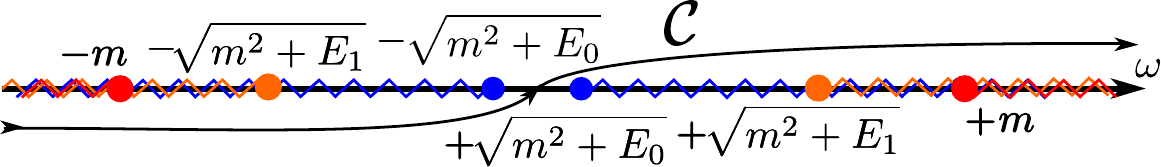}
\end{tabular}
\caption{\label{scalarbackp3} The integration contour  ${\cal C}$ in the complex $\omega$-plane. At the threshold when $E_{0}(U)\to -m^{2}$, the contour ${\cal C}$ is pinched by two branch points $\omega = \pm \sqrt{m^{2}+E_{0}}$.}
\end{figure}

If we consider a family of potentials controlled by some parameters, and tune these parameters in $U(x)$ to a certain threshold value, the effective action acquires an imaginary part.  In this case there is a simple way to argue that the threshold singularity will be of square root type and related to the lowest bound state in the potential $U(x)$. The effective action is proportional to the logarithm of the determinant of the Schr\"odinger operator 
\beq
i S_{\rm eff}(U) = -\frac{T}{2} \Tr \log \left(\partial_x^2+\omega^{2}-m^2-U(x)\right)\,, 
\eeq
where we used the Fourier representation for the time coordinate and, as we work in the approximation that the background is static, we obtain a factor of $T$ from the amount of time that the background has been switched on. Assuming that $E_{n}(U)$ is the  spectrum of the operator $-\partial_x^2+U(x)$, we find 
\beq\label{seffscdrap}
i S_{\rm eff}(U) = -\frac{T}{2} \sum_{n}\int_{{\cal C}} \frac{d\omega}{2\pi} \log \left(\omega^{2}-m^2-E_{n}(U)\right)\,,
\eeq
where  the index $n$ labels discrete  and continuous eigenstates. The contour ${\cal C}$ in the complex $\omega$-plane is chosen according to  the Feynman $i\epsilon$-prescription $m\to m-i\epsilon$, $\epsilon >0$. On the real axis we have multiple branch points $\omega = \pm \sqrt{m^{2}+E_{n}(U)}$, where we assume that the lowest bound state $E_{0}(U)>-m^{2}$ and the contour ${\cal C}$ goes above the branch cuts for $\omega >0$ and below the branch cuts for $\omega<0$ as shown in fig. \ref{scalarbackp3}. 
Because we assumed that $E_{0}(U)>-m^{2}$  we can Wick rotate the contour  ${\cal C}$ along the complex axis, so  $\omega\to i\omega$ and we obtain a manifestly real expression for the effective action
\beq
i S_{\rm eff}(U) = -i\frac{T}{2} \sum_{n}\int_{-\infty}^{+\infty} \frac{d\omega}{2\pi} \log \left(\omega^{2}+m^2+E_{n}(U)\right)\,.
\eeq
We see that the possibility to Wick rotate is related to the vacuum stability. When the lowest bound state $E_{0}(U)$ approaches $-m^{2}$ the branch points start pinching the contour ${\cal C}$, and this leads to the appearance of the singular terms in the effective action. So one can easily compute 
\beq \label{scseffres}
 S_{\rm eff}(U) = -\frac{T}{2} \sqrt{m^{2}+E_{0}(U)}+\dots\,,
\eeq
where we omitted less singular and non-singular terms. 

%The effective action is related to the vacuum energy; it sums all the zero point energy of the quantum harmonic oscillators sitting at each point in spacetime. The zero point energy of all harmonic oscillators is $m/2$ for $U(x)=0$. After switching on a nontrivial $U(x)$, we will have a new set of energy eigenstates with energy $\sqrt{m^2+E_n(U)}/2$, where $E_n(U)$ is a shift to the vacuum energy of the $n$-th oscillator. As we dial the shape of the potential, effectively decreasing the lowest energy eigenstate $E_{0}(U)$ , we expect  that at some point  $m^2+E_0(U)<0$, so the singularity in the effective action is proportional to $\sqrt{m^2+E_0(U)}$.

It is instructive to see how this singularity arises when we express the effective action through the scattering data related to the potential $U(x)$. Namely, below we are going to show  that the effective action can be expressed in terms of a logarithm of the transmission coefficient of a wave passing the potential $U(x)$. For the electric and gravitational cases this method will be more convenient.
%To understand physically what happens as we increase the value of $g$, we analyze classical equations of motion, using ``band theory".
%If we write the mode functions in terms of Fourier modes of frequency $\omega$, $\phi=\int \dd\omega e^{i\omega t}f(x)$, we see that they must solve the Schr\"odinger-like equation
%\beq\label{scalarschr}
%\left(\partial_x^2+(\omega^2-m^2-U(x))\right) f=0\,.
%\eeq
%The energy of excitations is given by
%\beq
%\omega_\pm=\pm \sqrt{p^2+m^2+U(x)} \, . 
%\eeq 
%We see that the band gap closes as we increase the size of $g$ (see fig. \ref{scalarbackp2}). As the band gap remains open far from the center of the potential, the above threshold excitations will be bound states than free particles. 
%\begin{figure}[h!]
%\centering
%\begin{tabular}{cc}
%\includegraphics[width=0.48\textwidth]{./figures/scalarp2.pdf}
%\end{tabular}
%\caption{\label{scalarbackp2} Here we show the ``bands" of the matter field. We expect a singularity in the effective action when the band gap closes, in the region where $U(x)$ is nonzero. We represent the narrowing of the band gap by the blue curves, in contrast with the red curves, where the band gap is unaltered. When the background is too strong, a bound state appears in the spectrum and it is energetically favourable to produce it. }
%\end{figure}

We begin by differentiating the effective action with respect to the mass, and obtain   
\beq\label{dereqsc}
i\frac{\partial S_{\rm eff}(U)}{\partial m^2}=-\frac{i}{2}\int \dd t\, \dd x\, G_F(t,x;t,x)\, , 
\eeq
where $G_F(t,x;t',x')\equiv \langle \phi(x,t)\phi(x',t')\rangle$ is the Feynman Green's function\footnote{We omit the time ordering symbol of the Feynman Green's function to avoid confusion with the time $T$ that the background is switched on.}. We can use the Fourier representation for the time components of the Green's function, which then satisfies
\beq\label{GFsc}
\left(\partial_x^2+\omega^{2}-m^2-U(x) \right) G_F(\omega;x,x')=i\delta(x-x')\, .
\eeq
By defining  mode functions $f_{\rm in}(x)$ and $f_{\rm out}(x)$, which are annihilated by the  Schr\"odinger operator 
\beq\label{deffifiout}
\left(\partial_x^2+\omega^{2}-m^2-U(x) \right)f_{\rm in/out}(x)=0\, 
\eeq
and satisfy the following boundary conditions
\bea\label{finout}
&f_{\rm in}(x)\xrightarrow{x\to-\infty}\displaystyle \frac{e^{-ipx}}{\sqrt{2p}}, ~~f_{\rm out}(x)\xrightarrow{x\to+\infty} \displaystyle \frac{e^{-ipx}}{\sqrt{2p}}\,, \quad p=\sqrt{\omega^2-m^2}\,,
\eea
we can express the Green's function as
\beq\label{grfsc}
G_F(\omega;x,x')=i\frac{f_{\rm in}(x)f^*_{\rm out}(x')\theta(x'-x)+(x\leftrightarrow x')}{W(f_{\rm in},f^*_{\rm out})} \, ,
\eeq
where the Wronskian is  $W(f_{\rm in},f^*_{\rm out}) \equiv f_{\rm in}(x)\partial_{x}f_{\rm out}^{*}(x)- f_{\rm out}^{*}(x)\partial_{x} f_{\rm in}(x)$. The functions $f_{\rm in}$ and $f_{\rm out}$ are related by Bogoliubov coefficients $\alpha$ and $\beta$
\bea
f_{\rm in}(x)=\displaystyle\alpha f_{\rm out}(x)+\beta f_{\rm out}^*(x)\,,
\eea
where $|\alpha|^{2}-|\beta|^{2}=1$, and a simple computation gives $W(f_{\rm in},f_{\rm out}^*)=i\alpha$.
We see that $1/\alpha$ is the transmission coefficient, which depends on $\omega, m$ and $U(x)$;
it can be obtained by solving the quantum mechanical scattering problem (\ref{deffifiout}). It is possible to show that the effective action $S_{\rm eff}(U)$ is controlled entirely by the coefficient $\alpha$ \cite{Nikishov:2002ez} (see also \cite{Polyakov:2007mm,Krotov:2010ma}).

In order to evaluate the effective action $S_{\rm eff}(U)$, we see from \eqref{dereqsc} and (\ref{grfsc}) that we must compute $\int \dd x\, f_{\rm in}(x) f_{\rm out}^*(x)$. It is possible to express this integral through the coefficient $\alpha$. For this we write the left-hand side of \eqref{deffifiout} for  $f_{\rm in}(x)$ with $m^2$, and multiply the equation by $f^*_{\rm out}(x)$ which is solution of the same equation but with $m^{2}+\delta m^{2}$. Analogously  we multiply the equation for $f^*_{\rm out}(x)$ with  $m^2+\delta m^2$ by $f_{\rm in}(x)$ with $m^2$. We subtract both expressions, integrate the result over $x$, and keep the first nontrivial terms in $\delta m^2$. This gives $\int \dd x f_{\rm in}(x) f^*_{\rm out}(x) = -i \partial \alpha /\partial m^{2}$, where we used the Feynman $i\epsilon$-prescription $m\to m-i\epsilon$, $\epsilon >0$.
We finally obtain
\beq\label{EAfsc}
S_{\rm eff}(U)=\frac{1}{2}iT \int_{\cal C} \frac{\dd \omega}{2\pi} \,  \log \alpha(\omega) \, , 
\eeq 
where  the choice of the contour ${\cal C}$ is explained above and shown  in figure \ref{scalarbackp3}. In appendix \ref{app:a}, we give a detailed derivation of this formula for the gravitational case, which is technically the most complicated.

As we see from \eqref{EAfsc}, finding $S_{\rm eff}(U)$ has now been reduced to a 1-D scattering problem.
 Again the singularity arises in the integral (\ref{EAfsc})  when the contour $\cal C$ is pinched by branch points.  
The result (\ref{EAfsc}) is not so surprising, as indeed in  the scattering theory  it is well-known that the transmission coefficient $1/\alpha$
is an analytic function of  energy $E$ on the physical sheet $\sqrt{E}$ ($\textrm{Im}\sqrt{E}>0$), except for  the points of  discrete spectrum $E=E_{n}$, in which the amplitude has  simple poles. Thus the coefficient $\alpha \sim (p-i\sqrt{|E_{0}|})/(p+i\sqrt{|E_{0}|})$ near the pinching branch points 
and computing the integral (\ref{EAfsc}) for $E_{0}(U) \to -m^2$  one recovers the result (\ref{scseffres}).
So we see that in the scattering approach the singularity mechanism is similar. More generally, the relation between scattering data and the determinant of the Schr\"odinger operator is well-known and has been thoroughly investigated \cite{Faddeev:1959yc}.

\subsection{Electric Fields}

\begin{figure}[h!]
\centering
\begin{tabular}{cc}
\includegraphics[width=0.48\textwidth]{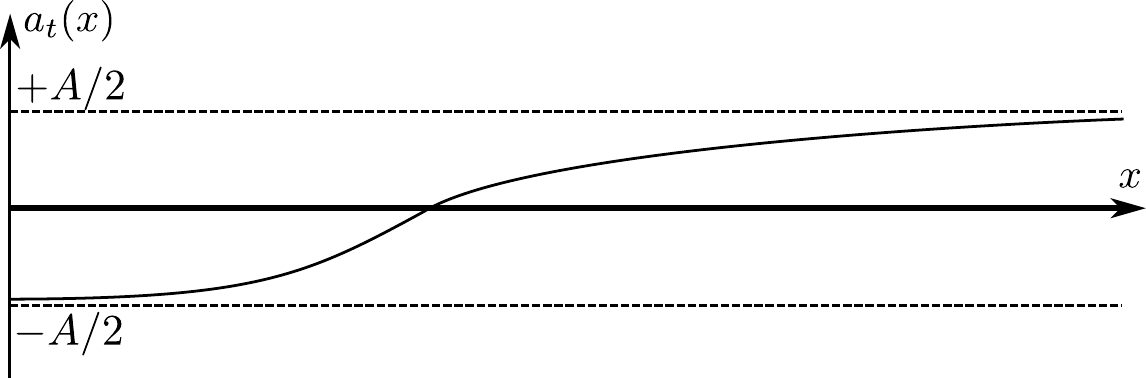}
\end{tabular}
\caption{\label{electricbranchesp1} Plot of a schematic form of the potential $a_t(x)$. We assume that $A<2m$.}
\end{figure}

Let us now consider the case of a free massive complex scalar in a strong electric field. We work in $1+1$ dimensions, but some of our results can be generalized to higher dimensions. The one-loop effective action is given by
\beq
\langle {\rm out} | {\rm in}\rangle = e^{i S_{\rm eff}(a)}=\int D\phi D\bar\phi \, \exp\left(i \int \dd t \dd x \left(|\partial_t\phi+i a_t\phi|^2-|\partial_x\phi|^2-m^2|\phi|^2\right)\right),
\eeq
where we picked the static gauge $a=a_t(x)\dd t$ for our background configuration, and chose $a_t(x)$ to be a smooth and monotonic function with asymptotic values $a_t(-\infty)=-A/2$ and $a_t(+\infty)=+ A/2$ (see fig. \ref{electricbranchesp1}).
The asymptotic values $\pm A/2$ are symmetric without loss of generality, by a simple shift of the potential. Other than monotonicity, we do not require the curve $a_t(x)$ to have any special property. We will see that particle production becomes favorable if $A> 2m$. If $A< 2m$, the effective action will be purely real and the vacuum is stable. As $ A\to 2m$ from below, we will show that the effective action acquires  a logarithmic singularity.

\begin{figure}[h!]
\centering
\begin{tabular}{cc}
\includegraphics[width=0.48\textwidth]{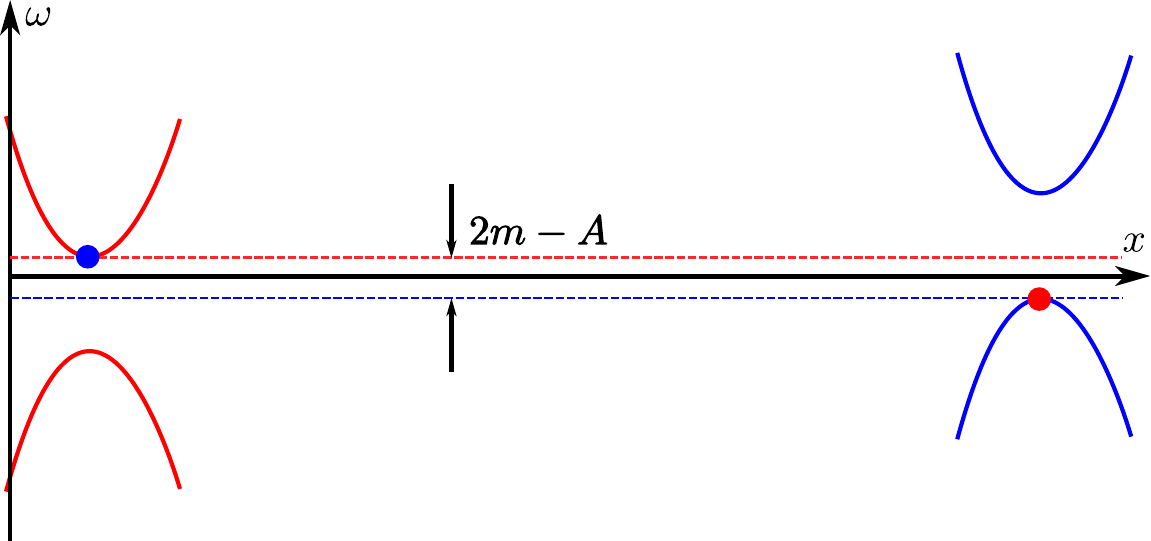}
\end{tabular}
\caption{\label{electricbranchesp2}  Plot of the ``bands" of the matter field. When the background is too strong the bottom of the valence band comes up to the top of the conduction band and it becomes energetically favorable to  trigger the tunneling and disrupt the vacuum by pair production.}
\end{figure}

To gain some intuition of the pair production threshold in the electric case, we analyze the classical equations of motion,  using band theory, in the asymptotic regimes $x\to \pm\infty$. The energies of excitations are given by
\beq
\omega_\pm=a_t(x)\pm \sqrt{p^2+m^2} \, . 
\eeq 
In figure \ref{electricbranchesp2}, we see that the maximum and minimum points of the energy  move as one goes from $x=-\infty$ to $x=+\infty$. When the bottom of the valence band comes up to the top of the conduction band, it becomes energetically favorable to disrupt the vacuum by pair production. This implies that threshold is reached when:
\beq
{\rm min\,} \left(\omega_+\right) -{\rm max\,}\left(\omega_-\right) = A-2m>0 \, . 
\eeq 

\begin{figure}[h!]
\centering
\begin{tabular}{cc}
\includegraphics[width=0.48\textwidth]{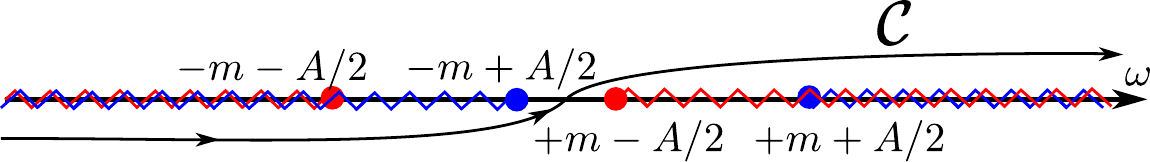}
\end{tabular}
\caption{\label{electricbranchesp3} The integration contour ${\cal C}$ in the complex $\omega$-plane.  The branch cuts here correspond to points where $p_{-}=p_{+}=0$. When the electric background near the threshold $A\to 2m$, the branch points $\omega =\pm(m-A/2)$ pinch the contour ${\cal C}$.}
\end{figure}

Now we proceed with the calculation of the effective action. Once again, it is convenient to differentiate it  by mass   
\beq\label{dereq}
i\frac{\partial S_{\rm eff}(a)}{\partial m^2}=-i\int \dd t\, \dd x\, G_F(t,x;t,x) , 
\eeq
where $G_F(t,x;t',x')\equiv \langle \phi^{*}(x,t)\phi(x',t')\rangle$ is the Feynman Green's function. We can use the Fourier representation for the time components of the Green's function, which then satisfies
\beq\label{GF}
\left(\partial_x^2+(\omega-a_t(x))^2-m^2 \right) G_F(\omega;x,x')=i\delta(x-x')\, .
\eeq
Finding $S_{\rm eff}(a)$ has now been reduced to a 1-D scattering problem similar to the scalar case we treated above. So we  define mode functions $f_{\rm in}$ and $f_{\rm out}$ which are annihilated by the operator in the left hand side of \eqref{GF}.   In terms of $f_{\rm in}$ and $f_{\rm out}$, the Green's function is given by
\beq
G_F(\omega;x,x')=i\frac{f_{\rm in}(x)f^*_{\rm out}(x')\theta(x'-x)+(x\leftrightarrow x')}{W(f_{\rm in},f^*_{\rm out})} \, . 
\eeq
The functions $f_{\rm in}$ and $f_{\rm out}$ satisfy the following boundary conditions
\bea\label{finout}
&f_{\rm in}(x)\xrightarrow{x\to-\infty}\displaystyle \frac{e^{-ip_-x}}{\sqrt{2p_-}}, ~~f_{\rm out}(x)\xrightarrow{x\to+\infty} \displaystyle \frac{e^{-ip_+x}}{\sqrt{2p_+}}\,,
\eea
where $p_{\pm}=\sqrt{(\omega\mp A/2)^2-m^2}$. The two solutions are related by Bogoliubov coefficients, $f_{\rm in}(x)=\alpha f_{\rm out}(x)+\beta f_{\rm out}^*(x)$. Using the same method as in the scalar case, we obtain for the effective action
\beq\label{EAf}
S_{\rm eff}(a)=iT \int_{\cal C} \frac{\dd \omega}{2\pi} \,  \log \alpha(\omega) \, .
\eeq 
The contour ${\cal C}$ must be chosen according to the Feynman $i\epsilon$-prescription $m\to m-i\epsilon$, $\epsilon>0$,
which gives $p_{\pm}\to p_{\pm}+i\epsilon$. There are multiple branch cuts on the real $\omega$-axis. They start at the points corresponding to zeros of  $p_{-}$ and $p_{+}$ and also $\alpha$. Therefore the contour ${\cal C}$ should go below the branch cuts for $\omega\to -\infty$ and 
 above the branch cuts for $\omega \to +\infty$ and pass between the left and right branch cuts near $\omega=0$ 
 (see fig. \ref{electricbranchesp3}). In general we may have a branch point which corresponds to $\alpha=0$ but when $A$ is very 
 close to $2m$ the branch cuts corresponding to $p_{-}=p_{+}=0$ will pinch the contour ${\cal C}$ first. We see that this mechanism is different from the scalar case, where the effect is due to  branch cuts corresponding to $\alpha=0$.

 So as $A\to2m$  the threshold is reached when the contour $\cal C$ is pinched by the branch points at $\omega=+m-A/2$ and  $\omega=-m+A/2$. This already hints at some universality, meaning that the most singular piece in the effective action will be largely agnostic about the particular shape of $a_t(x)$, but only depend on how close to threshold its maximum value is. Since the singularity appears as the branch points pinch the contour $\cal C$, we need to look at $\alpha(\omega\approx \pm(m-A/2))$. As the band gap closes, we can use an argument which gives a general form of the coefficient $\alpha$.  Leaving the details to appendix \ref{app:b}, when $\omega \approx \pm(m-A/2)$ and $A\approx 2m$,  the  coefficient $\alpha$ is given by an infinite series in small $p_{+}$ and $p_{-}$  and has the following form
\beq\label{alpapro}
\alpha=\frac{-ic_0 +c_- p_-+c_+ p_++ ic_{+-}p_{+}p_{-}+\dots}{2\sqrt{p_+p_-}}\, , 
\eeq
with $p_\pm\equiv\sqrt{(\omega\mp A/2)^2-m^2}$, and $(c_0, c_-, c_+,c_{+-})$ being shape-of-$a_t(x)$-dependent, but mass and frequency-independent real  numbers. Other coefficients in this expansion are not important for the singular terms in the effective action. The conservation of current implies $c_{+}c_{-}-c_{0}c_{+-}=1$.  

%Also, notice that the structure of the Bogoliubov coefficient is the same as the one in the quantum mechanics of a particle in a step function + delta function potential, with the size of the step and the strength of the delta function controlling the coefficients $c_0, c_-, c_+$. Physically, this means that at the threshold of vacuum instability, the dominant decay channel is that of a very soft pair. This pair has access to very coarse-grained features of the external potential $-$ the first few coefficients $c_0, c_-, c_+, c_{\pm}$ in the Taylor expansion for $\alpha$, and we can essentially approximate the problem by using the truncated form of $\alpha$ around small $p_+$ and $p_-$.

Having an expression for $\alpha$, we can evaluate the effective action. It is convenient to differentiate $\log \alpha(\omega)$ once by $m^{2}$, thus obtaining
\begin{align}
\frac{\partial\log \alpha(\omega)}{\partial m^{2}}=&-\frac{1}{2}\left(\frac{c_{+}+ic_{+-}p_{-}}{p_{+}}+\frac{c_-+ic_{+-} p_{+}}{p_{-}}+\dots\right)
 \nonumber\\&\times\frac{1}{-ic_0 +c_- p_{-}+c_+ p_{+}+\dots}+\frac{1}{4}\left(\frac{1}{p_{-}^{2}}+\frac{1}{p_{+}^{2}}\right).\label{mlog}
\end{align}
At this point the integral $\int_{\cal C} \dd\omega \frac{\partial}{\partial m^{2}}\log \alpha(\omega) $ is convergent and well defined. The last term in the right-hand side of (\ref{mlog}) has no branch cuts and can be evaluated in closed form; it is an uninteresting, non-singular piece of $S_{\rm eff}(a)$. So let us consider the first term in  (\ref{mlog}). We expect to obtain singular terms from vicinity of the points $\omega \approx \pm(m-A/2)$. 
%At threshold, $p_{\pm}\sim\sqrt{2m \omega}$, but the integral is not divergent at $\omega =0$ due to a nonzero $c_0$. 
%We can deform the contour in such a way that the integral is the sum of the piece above the branch point minus the integral over the piece %below the branch point; this integral runs from $m-A/2$ to some unimportant upper limit, say, $m$. It is only important that $m\gg m-A/2$. On this %contour, $p_{+} = i|p_{+}|$ and   $p_{-}=|p_{-}|$ and $p_{-}=-|p_{-}|$ above and below the cut correspondingly.   
It is possible to extract the non-analytic part from various integrals contributing to $\partial S_{\textrm{eff}}(a) /\partial m^{2}$. For instance it is not difficult to show that 
\begin{align}
&\int_{m-A/2}^{2m}\frac{d\omega}{p_{-}(-ic_{0}+c_{+}p_{+}+c_{-}p_{-}+ic_{+-}p_{+}p_{-}+\dots)} =\notag\\
&\qquad\qquad=k_{0}-\frac{c_{+}}{c_{0}^{2}}(2m-A) \log \left(\frac{2m-A}{2m}\right)+k_{1}(2m-A) +\dots\,, \label{intexample}
\end{align}
where the coefficients $k_{0}$ and $k_{1}$ depend  on $c_{0},c_{+},c_{-},c_{+-},\dots$ and on the upper limit of the integral, but the singular term depends only on $c_{0}$ and $c_{+}$. Analyzing various types of integrals arising from (\ref{mlog}) and similar to (\ref{intexample}) we finally obtain  the  most-singular non-analytic term of the effective action
\begin{align}
&\frac{\partial S_{\textrm{eff}}(a)}{\partial m^{2}} =  -\frac{T}{2\pi}\frac{c_{+} c_{-} - c_{0}c_{+-}}{2c_0^{2}}\left(\frac{2m-A}{2m}\right)\log \left(\frac{2m-A}{2m}\right)+\dots\, , 
\end{align}
and so it follows that
\begin{align}\label{singel}
S_{\textrm{eff}}(a) = \frac{T m^3}{2\pi c_0^{2}} \left(\frac{2m-A}{2m}\right)^{2}\log\left(\frac{2m-A}{2m}\right)+\dots\,,
\end{align}
where we have omitted  less singular  and non-singular terms.

Let us make a few comments about \eqref{singel}:
\begin{itemize}
\item The term $1/c_{0}^2$ is proportional to the transmission amplitude of the effective potential, thus for long smooth gauge fields it is exponentially damped. 
\item This term in the effective action is neither local in space (as in the usual derivative expansion) or in momentum space (as in the Euler-Heisenberg effective action). Neither of these representations can capture the threshold singularity, as we are always below threshold in the former case, and always above threshold in the latter. 
\item Despite depending on $A$, the effective action is gauge invariant, as $A=\int_{-\infty}^{+\infty}\dd x\, E(x)$. 
\item ${\rm Im}\,S_{\rm eff}$ can be reliably obtained by analytic continuation from \eqref{singel}, once we go slightly above the threshold, with $(A-2m)\ll m$. The amount of phase space available to pair produce depends on the dimension of the spacetime. A quick estimate gives
\beq
{\rm Im} \,S_{\rm eff}(a) \sim \int_0^{k_{\rm max}} \dd^{d-2} k \,\left(A-2\, m_{\rm eff}(k)\right)^2 \sim (A-2m)^\frac{d+2}{2}\, ,
\eeq 
where $m_{\rm eff}(k)\equiv \sqrt{m^2+k^2}$ is the effective mass of the produced particles, and the integral over transverse momenta runs over a finite range, determined by the condition $A-2m_{\rm eff}(k_{\rm max})=0$. For $d=4$, ${\rm Im}\, S_{\rm eff}(a) \sim (A-2m)^3$, as argued in \cite{Gies:2016coz}. Notice that $S_{\rm eff}(a)$ will contain a factor of $V_{d-2}$, the volume of the transverse directions, in higher dimensions. 
\item The expression \eqref{singel} is clearly invalid if $c_0=0$. In the regime $c_{0}\ll  (2m-A) \ll m$ one finds a different type of singularity 
\bea
S_{\rm eff}(a) =  -\frac{Tm}{2\pi} \frac{c_{+}c_{-}}{c_{+}^{2}+c_{-}^{2}} \left(\frac{2m-A}{2m}\right)\log\left(\frac{2m-A}{2m}\right)+\dots \,.
\eea 
We recover the expression above for    a ``quenching" electric field, with $a_t(x)=A/2\,\textrm{sgn}(x)$, where one has exactly $c_0=0$. 
\end{itemize}

As a particular example of the formulas presented above, we can determine the precise form of the Bogoliubov coefficient for a family of potentials $a_t(x)=A/2\tanh(x/l)$, parametrized by the width of the potential $l$. The result is presented in appendix \ref{app:c}. This family of potentials includes the quenching example when $l\to 0$, and we can show that the singularity is different in that case, confirming our last bullet point above. Expanding the Bogoliubov coefficient around small $p_+$ and $p_-$, we find the same structure argued for in this section, given by \eqref{alpapro}.

\subsection{Gravitational Fields}
\begin{figure}[h!]
\centering
\begin{tabular}{cc}
\includegraphics[width=0.48\textwidth]{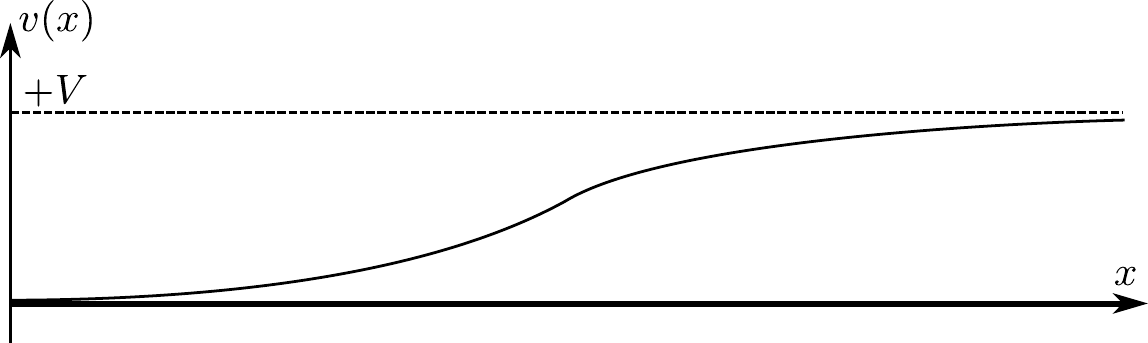}
\end{tabular}
\caption{\label{gravbranchesp1}  Plot of a schematic form of the gravitational potential $v(x)$. We assume that $V<1$.}
\end{figure}

Finally we consider the case of a massive free scalar field in a strong gravitational background. Again we work in $1+1$ dimensions and 
we would like  to determine the one-loop effective action
\begin{align}
\langle {\rm out} | {\rm in}\rangle = e^{iS_{\rm eff}(v)} = \int D\phi \, \exp\left(-\frac{i}{2}\int \dd t \dd x \sqrt{-g} \left(g^{\mu\nu}\partial_{\mu}\phi\partial_{\nu}\phi+m^{2}\phi^{2}\right)\right)\,.
\end{align}
For our purposes it is convenient to describe the metric $g_{\mu\nu}$ in Painlev\'e-Gullstrand coordinates,
\beq\label{PG}
ds^2=g_{\mu\nu}dx^{\mu}dx^{\nu}=dt^2-(dx-v(x)dt)^2\,,
\eeq 
where we choose $v(-\infty)=0$ and let $v(x)$ increase smoothly for growing $x$ up to some value $v(+\infty)=V$, similarly to the electrostatic potential $a_{t}(x)$  (see figure \ref{gravbranchesp1}). 
This geometry would  have a horizon at $x_h$ if   $v(x_h)=1$ and  thus $g_{tt}=0$. As we will show later, the vanishing of $g_{tt}$ coincides with the criterion for vacuum decay. For $x$ at which $v(x)<1$, we can interpret $v(x)$ as the escape velocity from the position $x$~\cite{Hamilton:2004au}. So our criterion for vacuum stability is that $v(x)<1$ for all $x$. Therefore we assume that $V<1$ but we tune $V$ to the threshold value, i.e. $V\to1$. This case is mathematically closer to the equipotential planes with fixed asymptotics, in the electric case of the previous section. We will show that the effective action acquires a square root singularity when $V\to1$.

\begin{figure}[h!]
\centering
\begin{tabular}{cc}
\includegraphics[width=0.48\textwidth]{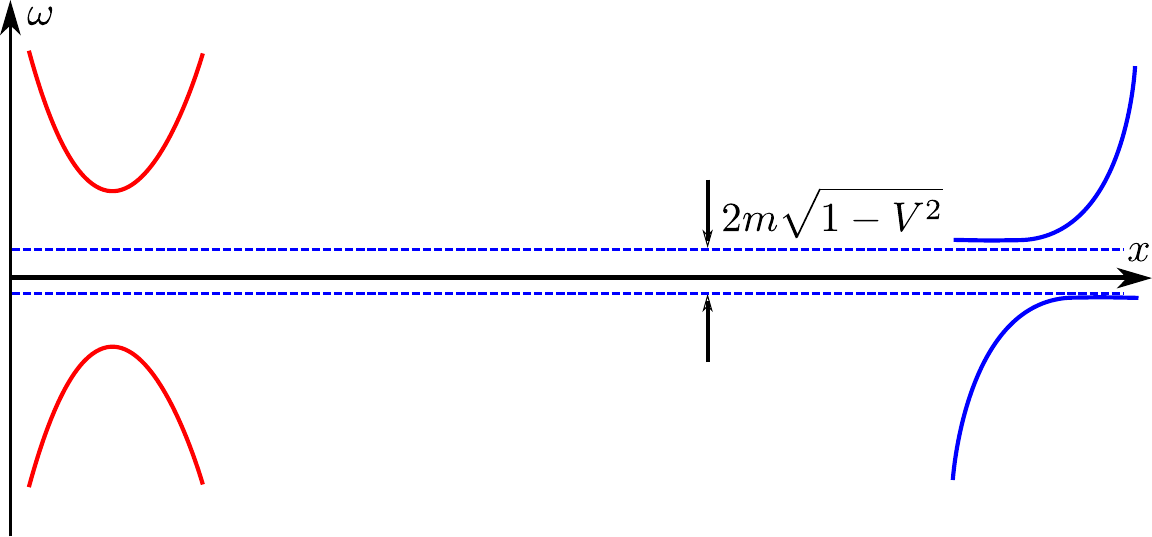}
\end{tabular}
\caption{\label{gravbranchesp2} Here we show the bands and the dominant tunneling event, which comes from the top of the lower blue band touching the bottom of the upper blue band.}
\end{figure}

Let us  consider the semiclassical analysis for the gravitational case. The nature of the gap is slightly different than in the electric case. This is due to the different structure of the single-particle Hamiltonian \cite{PiTPlec,Volovik:2008ww,Volovik:2009eb}. The energies of excitations are given by
\beq
\omega_\pm(p,x) = p \, v(x) \pm \sqrt{p^2+m^2}\, ,
\eeq and threshold corresponds to $V\to1$, as 
\beq 
{\rm min} \left(\omega_+\right) -{\rm max}\left(\omega_-\right) = 2m \sqrt{1-V^2} \, .
\eeq 
The purpose of the mass term is just to open a gap between positive and negative energy bands (see fig. \ref{gravbranchesp2}), as particle production can occur for gravitational fields without a horizon when the matter sector is gapless~\cite{Pimentel:2015iiv}. 

Notice that when $v(x)=1$, $g_{tt}=0$, which, in these coordinates, is the usual definition of the horizon. In other words, our criterion for the location of the horizon being the distance at which it becomes energetically favorable to pair produce coincides with the definition coming from the causal structure of spacetime. Also, notice that a shift $v\to v+C$ is not unphysical, for the case of a static metric. In order to remove the constant $C$, one could use a Galilean transformation, but this would imply a redefinition of the time coordinate. In other words, the criterion for vacuum instability depends not just on the difference between the asymptotic values of the velocity field, like in the electric field example, but also on its absolute values as $x\to \pm \infty$.

As usual to compute the effective action we differentiate it  by the mass term  
\begin{align}
i\frac{\partial S_{\rm eff}(v)}{ \partial m^{2}} = -\frac{i}{2} \int \dd t\, \dd x\,  G_{F}(x,t;x,t)\,,
\end{align}
where $G_{F}(x,t;x',t')\equiv\langle \phi(x,t)\phi(x',t')\rangle$ is the Feynman Green's function and we used that for our metric $\sqrt{-g}=1$. The Green's function obeys the equation 
\begin{align}
\left(\partial_{x}((1-v^{2})\partial_{x}) -2 v \partial^{2}_{xt}-(\partial_{x}v)\partial_{t}-\partial_{t}^{2} -m^{2}\right)G_{F}(x,t;x',t') = i \delta(x-x')\delta(t-t')\,.\label{Fgrfgr}
\end{align}
%where the operator $L_{x,t}$ reads
%\begin{align}
%L_{x,t} = \partial_{x}((1-v^{2})\partial_{x}) -2 v \partial^{2}_{xt}-v'\partial_{t}-\partial_{t}^{2} -m^{2}\,. \label{grKG}
%\end{align}
This equation contains a term with a first order derivative in $x$, which naively makes it difficult to apply the previous strategy of expressing  the effective action as an integral over the logarithm of the transmission coefficient. Nevertheless, by properly changing variables, we are able to obtain a similar $\log \alpha$ formula for the effective action. 

%To proceed we make the following change of functions  
%\begin{align}
%\phi(x,t) = \frac{e^{i\omega t} e^{i \chi(x)}}{\sqrt{1-v^{2}}} f_\omega(x), \quad \partial_{x}\chi(x) = \frac{\omega v(x)}{1-v^{2}(x)}\,,
%\end{align}
%then one has
%\begin{align}
%L_{x,t}\phi(x,t) = \frac{e^{i\omega t} e^{i\chi(x)}}{\sqrt{1-v^{2}}} (1-v^{2})\mathcal{L}_{\omega,x}f_\omega(x)\,,
%\end{align}
%where the operator $\mathcal{L}_{\omega,x}$ is of a Schr\"odinger-like type
%\begin{align}
%\mathcal{L}_{\omega,x}= \partial_{x}^{2} + \frac{\omega^2-m^2(1-v^2)+(\partial_x v)^2}{(1-v^2)^2}+\frac{v \, \partial_x^2 v}{1-v^2}\,. \label{SchropGr}
%\end{align}

To proceed one can check that the  Green's function $G_{F}(x,t;x',t')$ can be written in the form 
\begin{align}
G_{F}(x,t;x',t') = \int \frac{\dd\omega}{2\pi}\frac{e^{i\omega(t-t')}e^{i(\chi(x)-\chi(x'))}}{\sqrt{(1-v^{2}(x))(1-v^{2}(x'))}}G_{\omega}(x,x')\,, \label{grgranz}
\end{align}
where the new Green's function $G_{\omega}(x,x')$ obeys a Schr\"odinger-like equation
\begin{align}\label{ngrfgr}
\left(\partial_{x}^{2} + \frac{\omega^2-m^2(1-v^2)+(\partial_x v)^2}{(1-v^2)^2}+\frac{v \, \partial_x^2 v}{1-v^2}\right)G_{\omega}(x,x')= i\delta(x-x')\,.
\end{align}
and the function $\chi$ is defined as $\partial_{x}\chi =\omega v/(1-v^{2})$. 
It is easy to check that (\ref{grgranz}) indeed satisfies (\ref{Fgrfgr}). Therefore for the effective action we obtain
\begin{align}
\frac{\partial S_{\rm eff}(v)}{ \partial m^{2}} &=-\frac{T}{2}  \int \frac{\dd\omega}{2\pi} \int \dd x  \frac{G_{\omega}(x,x)}{1-v^{2}(x)}\,. \label{grefact2}
\end{align}
The Green's function $G_{\omega}(x,x')$ as usual can be expressed through $f_{\rm in}$ and $f_{\rm out}$  functions
\begin{align}
G_{\omega}(x,x') = i\frac{f_{\textrm{in}}(x) 
f_{\textrm{out}}^{*}(x')\theta(x'-x)+(x\leftrightarrow x')
}{W(f_{\textrm{in}},f^{*}_{\textrm{out}})}\,, \label{Ggrthf}
\end{align}
where $f_{\rm in}$ and $f_{\rm out}$ are anihilated by the Schr\"odinger-like operator in (\ref{ngrfgr})   and have  asymptotics
\begin{align}
&f_{\rm in}(x)  \xrightarrow{x\to-\infty}\displaystyle \frac{1}{\sqrt{2p_{-}}}e^{-ip_{-}x}, \quad f_{\rm out}(x)  \xrightarrow{x\to+\infty}\displaystyle \frac{1}{\sqrt{2p_{+}}}e^{-ip_{+}x}\,, \label{asexp}
\end{align}
where we denoted
\begin{align}
p_{-}= \sqrt{\omega^{2}-m^{2}}, \quad p_{+} =\frac{1}{1-V^{2}}\sqrt{\omega^{2}-m^{2}(1-V^{2})}\,,
\end{align}
and, as usual, we define $\alpha$ and $\beta$ as $f_{\rm in}(x) =\displaystyle \alpha f_{\rm out}(x)+\beta f_{\rm out}^{*}(x)$.
As shown in appendix \ref{app:a}  we can  bring the formula (\ref{grefact2}) to our usual form
\begin{align}
S_{\rm eff}(v) =  \frac{1}{2}iT  \int_{\cal C} \frac{d\omega}{2\pi}\log \alpha(\omega)\,. \label{grlogaf}
\end{align}

\begin{figure}[h!]
\centering
\begin{tabular}{cc}
\includegraphics[width=0.48\textwidth]{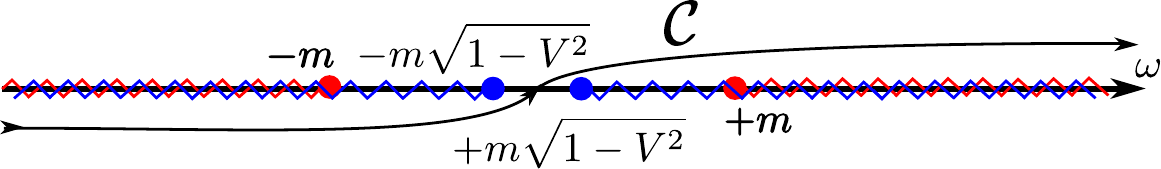}
\end{tabular}
\caption{\label{gravbranchesp3} The integration contour ${\cal C}$ in the complex $\omega$-plane. Branch points $\omega =\pm m \sqrt{1-V^{2}}$ corresponding to $p_{+}=0$ pinch the contour ${\cal C}$ near $\omega \approx 0$ when $V \to1$. This ``pinching'' region determines the singular piece of the effective action.}
\end{figure}

The pinching singularity comes again from very small frequencies near the points where $p_{+}=0$, and we need to determine $\alpha$ for $\omega \sim m \sqrt{1-V^2}$  (see figure \ref{gravbranchesp3}). 
Notice that the relevant feature is the behavior of the mode functions as $x\to\infty$, and for $\omega \sim m \sqrt{1-V^2}$, the mode function has very small modulation with $\omega$. This means that the transmission coefficient behaves like
\beq
\alpha\approx\frac{-i d_0+ d_+ p_+ +\dots}{2\sqrt{p_+}}\, , 
\eeq
as $p_-\approx m$, and the model dependence is encoded in the coefficients $(d_0, d_+)$. We can once again take derivatives of the effective action to isolate its singular piece. It turns out that differentiating once with respect to $m^2$ is enough to isolate the singular term. Following the same steps as in the electric case, we arrive at 
\beq\label{gravres}
S_{\rm eff}(v) =  m T \sqrt{1-V^2} +\dots \, .
\eeq
In the appendix \ref{app:c} we find an exact $\alpha(\omega)$ for the step potential $v(x)=V\theta(x)$. In this case one can calculate the integral over $\omega$ in  (\ref{grlogaf}) exactly and obtain the result (\ref{gravres}).

This singular part of the effective action can be written in a ``local" form\footnote{The volume form on the timelike surface $x\to +\infty$ is given by $\sqrt{1-V^2}=\sqrt {g_{\rm ind}}$.} due to the different tunneling pattern when the vacuum breaks down -- pairs are produced at large $x$, rather than at both small and large $x$. There is also an analogous term to the electric threshold result, $\sim (1-V^2)\log (1-V^2)$, but it is subleading to \eqref{gravres}. Notice that even in the special case $d_0=0$ we obtain the same singularity, albeit with different overall coefficient. Another interesting thing is that the leading term \eqref{gravres} does not care about the detailed coefficients $d_{0,+}$ $-$ as long as they are nonzero, the only relevant data from the metric is the value of $V$. This is unlike the electric case, where the leading singular term depends on $2m-A$ but also on $c_0$.

This threshold singularity is a quantum analog of Choptuik scaling. Choptuik considered a family of initial data labeled by a parameter $p$. Under time evolution using Einstein's equations, he found~\cite{Choptuik:1992jv} that the final state had a black hole of mass $M\sim (p-p_{\rm cr})^\gamma$, for $p>p_{\rm cr}$. The exponent $\gamma$ is largely independent on the details of the family of initial data. 

Above criticality, the formation of a black hole indicates the appearance of a horizon. Our critical exponent is entirely analogous, but is a quantum diagnostic of the appearance of the horizon. In our case, we look at $S_{\rm eff}(v)$ rather than $M$, criticality is reached when $V=1$, and the critical exponent is $\gamma=1/2$ \footnote{Interestingly, an analysis of black hole formation in a different context gives the same critical exponent~\cite{Strominger:1993tt}.}.

\section{Conclusions} \label{s:conclusions}
In this paper, we argued that the crossover between the quantum mechanical stability and instability of background fields has certain universal features. This is largely due to the first unstable process triggered right above threshold having very long wavelength and low energy. This soft emission process probes only the roughest features of the external background, and the threshold singularity can be easily expressed in terms of rough background data. There are many avenues for further investigation:
\begin{itemize}
\item Our analysis was restricted to gaussian matter fields. How would interactions in the matter sector change the critical exponents in the threshold singularity? 
\item Can we connect our results to existing methods for treating backreaction in black holes~\cite{Volovik:1999fc,Parikh:1999mf}? It would be nice to incorporate our threshold singularity to the problem of formation of a black hole, in order to see if vacuum polarization delays its formation, or prevents formation whatsoever for initial data close enough to threshold. 
\item Finally, it would also be interesting to find the threshold singularity for more realistic field configurations: for example, a spherically symmetric configuration, like a star, where we take a mass shell to be very close to its Schwarzschild radius. 
\end{itemize}
We leave such fascinating problems to the near future.

%%%%%%%%%%%%%%%%%%%%%%%%%%%%%%%%%%%%%%%%

\vskip 5 pt

\paragraph{Acknowledgements} We would like to thank Daniel Baumann, Garrett Goon, Diego Hofman, Viatcheslav Mukhanov, Andrew Strominger, Leon Takhtajan and Grigory Volovik for useful discussions, and Daniel Baumann, Garrett Goon and Diego Hofman for comments on a draft. The work of G.T. was supported  by the MURI grant W911NF-14-1-0003 from ARO and by DOE grant de-sc0007870. G.P. acknowledges funding from the European Union's Horizon 2020 research and innovation programme under the Marie-Sk\l{}odowska Curie grant agreement number 751778. The work of G.P. is part of the Delta ITP consortium, a program of the Netherlands Organisation for Scientific Research (NWO) that is funded by the Dutch Ministry of Education, Culture and Science (OCW). G.P. also acknowledges support from a Starting Grant of the European Research Council (ERC STG Grant 279617). 

\appendix
\section{Derivation of $S_{\rm eff}=i\int d\omega \log\alpha$}\label{app:a}

In this appendix, we derive the formula for the gravitational effective action in terms of the Bogoliubov coefficient $\alpha$ in detail. The electromagnetic and scalar cases simply follow from this.

%We sketched the derivation in the main text in words, but provide all nontrivial steps below. We work out the cases of gravity and electrodynamics as examples, but extension to other setups is straightforward. The case of gravity is slightly more complicated due to the Klein-Gordon operator not being of the Schr\"odinger form, so we proceed with it more carefully. With the electrodynamics case we omit some of the details.

Using the expression (\ref{Ggrthf}) for $G_\omega(x,x')$  and formula (\ref{grefact2}) we obtain for the effective action
\begin{align}
\frac{\partial S_{\rm eff}(v)}{\partial m^{2}} =-i\frac{T}{2}  \int \frac{\dd\omega}{2\pi} \int \dd x  \frac{f_{\rm in}(x)f_{\rm out}^{*}(x)}{(1-v^{2}(x))W(f_{\rm in},f^{*}_{\rm out})}\,, \label{intoverx}
\end{align}
where one can easily calculate $W(f_{\rm in},f^{*}_{\rm out})=i\alpha$.  We must now calculate the integral over $x$ in (\ref{intoverx}). In order to proceed, we do the following: consider the equations 
\begin{align}
 &\partial_{x}^{2}f_{{\rm in},m^{2}}(x) + \left(\frac{\omega^2-m^2(1-v^2)+(\partial_x v)^2}{(1-v^2)^2}+\frac{v \, \partial_x^2 v}{1-v^2}\right)f_{{\rm in},m^{2}}(x)=0\,,\notag\\
 &\partial_{x}^{2}f^{*}_{{\rm out},m^{2}+\delta m^{2}}(x) + \left(\frac{\omega^2-(m^2+\delta m^2)(1-v^2)+(\partial_x v)^2}{(1-v^2)^2}+\frac{v \, \partial_x^2 v}{1-v^2}\right)f^{*}_{{\rm out},m^{2}+\delta m^{2}}(x)=0\,.
\end{align}
Multiplying the first equation by $f^{*}_{{\rm out},m^{2}+\delta m^{2}}(x)$ and the second by $f_{{\rm in},m^{2}}(x)$ and subtracting them we obtain 
\begin{align}
\partial_{x}(f_{{\rm in},m^{2}}\partial_{x}f^{*}_{{\rm out},m^{2}+\delta m^{2}}-f^{*}_{{\rm out},m^{2}+\delta m^{2}}\partial_{x}f_{{\rm in},m^{2}}) = \delta m^{2} \frac{f_{{\rm in},m^{2}}(x)f^{*}_{{\rm out},m^{2}+\delta m^{2}}(x)}{1-v^{2}(x)}\,.
\end{align}
Integrating over $x$ the left and the right parts  from $-L$ to $L$, where $L\to +\infty$ we get 
\begin{align}
\delta m^{2} \int_{-L}^{+L} \dd x \frac{f_{{\rm in},m^{2}}(x)f^{*}_{{\rm out},m^{2}+\delta m^{2}}(x)}{1-v^{2}(x)}
 = (f_{{\rm in},m^{2}}\partial_{x}f^{*}_{{\rm out},m^{2}+\delta m^{2}}-f^{*}_{{\rm out},m^{2}+\delta m^{2}}\partial_{x}f_{{\rm in},m^{2}})|_{-L}^{+L}\,.  
 \end{align}
Because we take $L\to \infty$, we can use the asymptotic expressions for $f_{\rm in,out}$ (\ref{asexp}) and find 
\begin{align}
&(f_{{\rm in},m^{2}}\partial_{x}f^{*}_{{\rm out},m^{2}+\delta m^{2}}-f^{*}_{{\rm out},m^{2}+\delta m^{2}}\partial_{x}f_{{\rm in},m^{2}})|_{-L}^{+L}   =\notag\\
&=\delta m^{2}\left(-i\frac{\partial \alpha}{\partial m^{2}}-L \alpha \frac{\partial (p_{-}+p_{+})}{\partial m^{2}}  -\frac{1}{2}i\Big(\beta^{*}\frac{\partial \log p_{-}}{\partial m^{2}}e^{2iLp_{-}}-\beta\frac{\partial\log p_{+}}{\partial m^{2}}e^{2iLp_{+}}\Big)\right)+\dots\,.
 \end{align}
We use the Feynman $i\epsilon$-prescription $p_{\pm}\to p_{\pm}+i\epsilon$ with infinitesimal $\epsilon>0$, so the oscillating terms above are zero for large $L$. Finally we obtain
\begin{align}
&(f_{{\rm in},m^{2}}\partial_{x}f^{*}_{{\rm out},m^{2}+\delta m^{2}}-f^{*}_{{\rm out},m^{2}+\delta m^{2}}\partial_{x}f_{{\rm in},m^{2}})|_{-L}^{+L}   =-i\delta m^{2}\alpha \frac{\partial}{\partial m^{2}}\Big(\log \alpha +L (p_{-}+p_{+})\Big)\,.
 \end{align}
Putting this together, we  find 
\begin{align}
\frac{\partial S_{\rm eff}(v)}{\partial m^{2}} =i\frac{T}{2}  \int_{\cal C} \frac{\dd\omega}{2\pi}\frac{\partial}{\partial m^{2}}\Big(\log \alpha +L (p_{-}+p_{+})\Big).
\end{align}
Our last task is to argue that the terms proportional to $L$ are unimportant. By that we mean that they only carry uninteresting dependence on the background. The term proportional to $L p_-$ is harmless, depending only on $m$, but the term proportional to $Lp_+$ seems to have nontrivial dependence on $V$. Let us write it more explicitly
\beq
\int_{\cal C} \dd\omega\, L p_{+}=\int_{\cal C} \dd\omega \frac{1}{1-V^{2}}(\omega^{2}-m^{2}(1-V^{2}))^{1/2}\,.
\eeq
If we change variables $\omega= \omega'(1-V^2)^{1/2}$ then the $V$ dependence drops out of the integral and it is exactly equal to the $L p_-$ integral. This argument is too fast, as the integral is UV divergent. The correct argument is that the cutoff is background dependent. For the $p_-$ integral, we are at $x\to-\infty$ so we choose some hard cutoff $\Lambda$ in frequency space. At $x\to+\infty$, the metric is $\dd t^2 (1-V^2)$ so we must choose the cutoff $\Lambda/(1-V^2)^{1/2}$ in frequency space, to take into account the warping of time intervals. This renders the $Lp_+$ integral to have no interesting dependence on $V$. In summary, up to non-important terms, we find 
\begin{align}
S_{\rm eff}(v) =  \frac{1}{2}iT  \int_{\cal C} \frac{d\omega}{2\pi}\log \alpha(\omega)\,.
\end{align}

\section{Behavior of $\alpha$ Near Threshold}\label{app:b}
In this appendix we derive the behavior of the Bogoliubov coefficient $\alpha$ when the effective mass gap is very small. In other words, we find the first few terms in an expansion for $\alpha$ around vanishing mass gap. To start, let us consider the Schr\"odinger equation 
\begin{align}
(\partial_{x}^{2}+U(x)-m^2)f =0\,, \label{mq2}
\end{align}
where $U(x)$ either switches off or asymptotes to some fixed values $U(\pm \infty)$ in a smooth way. We are interested in the cases where the effective mass gap at infinity
\beq
p_{\pm}^2\equiv U(\pm \infty)-m^2
\eeq
is very small, namely $p_{\pm}^2\ll m^2$. We consider here the case in which both $p_{\pm}^2$ are small. In the main text, the gravitational background is such that only in one extreme the mass gap vanishes. Applying our formulas to that example is straightforward.

In the region outside of which $U(x)$ is varying, the mass term is either $p_+^2$ or $p_-^2$, which we assume are small. Neglecting those terms, we get 
\begin{align}
\partial_{x}^{2}f =0\,,
\end{align}
therefore the solutions of the equations of motion are 
\begin{align}
f = a_{1}+b_{1}x \quad x\ll 0, \qquad f = a_{2}+b_{2}x, \quad x\gg 0 \label{ap1}
\end{align}
where the potential varies significantly close to $x=0$. The coefficients $(a_{1},b_{1})$ and $(a_{2}, b_{2})$ are linearly dependent 
\begin{align}
a_{1}= c_- a_{2} + c_{+-} b_{2}, \quad b_{1}=c_0 a_{2}+ c_+ b_{2}\,, \label{con}
\end{align}
where the coefficients $(c_0, c_+, c_-, c_{+-})$ are independent of $p_{\pm}$ (as $p_{\pm}$ do not appear in the differential equation with linear functions as solutions), and, from the conservation of current, it follows that we can choose $(c_0, c_+, c_-, c_{+-})$ to be real, with 
$c_+ c_- - c_0 c_{+-} =1$. Then matching the solutions (\ref{ap1}) with the asymptotic solutions in terms of plane waves, we obtain 
\begin{align}
 a_{1} = \frac{1}{\sqrt{2p_{-}}}, \quad b_{1} = \frac{-ip_{-}}{\sqrt{2p_{-}}}, \quad a_{2} = \frac{\alpha+\beta}{\sqrt{2p_{+}}}, \quad b_{2}= \frac{-ip_{+}(\alpha-\beta)}{\sqrt{2p_{+}}}\,,
\end{align}
and solving the equations (\ref{con}) we get 
\begin{align}\label{expralph}
&\alpha =\frac{-i c_0 +c_- p_{-} +c_+ p_{+}+i c_{+-} p_{-}p_{+}}{2\sqrt{p_{-}p_{+}}}\,,\quad \beta =\frac{i c_0 -c_- p_{-} +c_+ p_{+}+i c_{+-} p_{-}p_{+}}{2\sqrt{p_{-}p_{+}}}\,.
\end{align}
Now having $\alpha$ we can evaluate the effective action. These expressions are only valid for $|p_\pm|\ll m$.

\section{Exact Solutions}\label{app:c} 

\subsection{Electric example}
The exact solution in the electric case is available for the gauge field profile $a_t(x) = A/2\tanh(x/l)$. In this case one can obtain an exact Bogoliubov coefficient $\alpha$; it is given by 
\beq\label{alphatanh}
\alpha = \frac{i/l}{\sqrt{p_{+}p_{-}}}\frac{\Gamma(1-ip_{-}l)\Gamma(1-ip_{+}l)}{\Gamma(\rho-\frac{i}{2}l(p_{-}+p_{+}))\Gamma(1-\rho-\frac{i}{2}l(p_{-}+p_{+}))}\,,
\eeq
where $\rho = \frac{1}{2}+\frac{1}{2}\sqrt{1-A^{2}l^{2}}$ and $p_{-},p_{+}$ are defined below (\ref{finout}). If we first tune $l\to 0$ we obtain the step potential $a_t(x)=A/2\, \textrm{sgn}(x)$ and the Bogoliubov coefficient (\ref{alphatanh}) simplifies to $\alpha = (p_{-}+p_{+})/2\sqrt{p_{+}p_{-}}$. On the other hand, if $l$ is fixed and we are in the regime where $p_{+}$ and $p_{-}$ are small, we find 
\beq
\alpha = \frac{\sin \pi \rho}{\pi l}\frac{(2i-l(\psi(\rho)+\psi(1-\rho)+2\gamma_E)(p_{-}+p_{+})+\dots)}{2\sqrt{p_{+}p_{-}}}\,, 
\eeq
where $\psi(x)$ is the digamma function, and $\gamma_E$ is the Euler-Mascheroni constant. This form of the $\alpha$ coefficient agrees with (\ref{expralph}).    
%%%%%%%%%%%%%%%%%%%%%%%%%%%%%%%%%%%%%%%%%%

\subsection{Gravitational example}
In this subsection we are going to find the coefficient $\alpha$ in the case of a step potential $v(x)=V\theta(x)$. 
To proceed it is convenient to write the Schr\"odinger equation  for the operator (\ref{ngrfgr}) as
\begin{align}
\partial_{x}\left((1-v^{2})\partial_{x}\Big(\frac{f_{\rm in}(x)}{\sqrt{1-v^{2}}}\Big)\right)+\frac{\omega^{2}-m^{2}(1-v^{2})}{(1-v^{2})^{3/2}}f_{\rm in}(x) =0\,. \label{she2}
\end{align}
Now integrating this equation from $x=-\delta$ to $x=\delta$ with $\delta \to 0$ we find the boundary conditions for $f_{\rm in}(x)$ and $\partial_{x}f_{\rm in}(x)$ at $x=0$:
\begin{align}
f_{\rm in}(0^{+}) = \sqrt{1-V^{2}}f_{\rm in}(0^{-}), \quad \sqrt{1-V^{2}}\partial_{x}f_{\rm in}(0^{+}) = \partial_{x}f_{\rm in}(0^{-})\,.
\end{align}
Using these boundary conditions one can find 
\begin{align}
\alpha = \frac{p_{-}+(1-V^{2})p_{+}}{2\sqrt{(1-V^{2})p_{-}p_{+}}}\,,
\end{align}
where $p_{-}=\sqrt{\omega^{2}-m^{2}}$ and $p_{+}=\sqrt{\omega^{2}-m^{2}(1-V^{2})}/(1-V^{2})$.

\newpage
\bibliographystyle{utphys}
\bibliography{MassPartProd}

\end{document}